\documentclass[preprint,12pt]{iopart}
\usepackage{epsfig,graphicx,amssymb,soul,color,lscape,times}
\begin{document}
%-------------------------------------------------------------------
\title[]{Thermal Radiation from Compact Objects in Curved Space-Time}
\author{Abhas Mitra$^1$, Krishna Kumar Singh$^{1,2}$}
\ead{abhasmitra@gmail.com,~~kksastro@barc.gov.in}
{}$^1$ {Homi Bhabha National Institute, Mumbai- 400 094, India}\\
{}$^2$ {Astrophysical Sciences Division, Bhabha Atomic Research Centre, Mumbai- 400 085, India}

%-----------------------Abstract------------ 
\begin{abstract}\\
We highlight  here  the  fact  that the distantly observed luminosity of a spherically symmetric compact star radiating thermal radiation isotropically 
is higher by a factor of $(1+z_{\rm b})^2$ compared to the corresponding flat space-time case, where $z_{\rm b}$ is the surface gravitational redshift 
of the compact star. In particular, we emphasize that if the thermal radiation is indeed emitted isotropically along the respective normal directions at 
each point, this factor of increment $(1+z_{\rm b})^2$  remains unchanged even if the compact object would lie within its {\em photon sphere}. Since a 
canonical neutron star has $z_{\rm b} \approx 0.1$, the actual X-ray luminosity from the neutron star surface could be $\sim 20 \%$ higher than what 
would be interpreted by ignoring the general relativistic effects described here. For a static compact object, supported by only isotropic pressure, 
compactness is limited by the Buchdahl limit $z_{\rm b} < 2.0$. However, for compact objects supported by anisotropic pressure, $z_{\rm b}$ could 
be even higher ($z_{\rm b} < 5.211$). In addition, in principle, there could be ultra-compact objects having $z_{\rm b} \gg 1$. Accordingly, the general 
relativistic effects described here might be quite important for studies of thermal radiation from some ultra-compact objects.\\
{\bf Keyword} relativistic astrophysics; gravitational redshift; neutron stars, X-ray luminosity
\end{abstract}
\maketitle
%----------------------------------------Section-1: Introduction------------------------------------
\section{Introduction}
Stellar remnants, such as white dwarfs (WDs), neutron stars (NSs) and black holes (BHs), mostly formed through the gravitational collapse of normal stars 
at the end of their life, are referred to as compact objects in astrophysics. It is well known that although general relativity may play a minor role 
for most of the observational astrophysics, it can become significant for studies of NSs and BH candidates (BHCs) or Ultra-Compact Objects (UCOs). 
A rough measure of the ``compactness'' of an astrophysical object may be found by the parameter $GM/R_{\rm b} c^2$, where $G$ is the constant of gravitation, 
$c$ is the speed of light, $M$ is the gravitational mass and $R_{\rm b}$ is the radius of the object. For~the Sun, the compactness is very small 
$\approx 2 \times 10^{-6}$. Since a WD can be as massive as the Sun and yet have a radius of $\sim$$1\%$ of the Sun, its compactness can be two orders higher $\sim$$2 \times 10^{-4}$. On~the other hand, it is usually believed that for an NS, this compactness is $\sim$$0.1$. 
\par
The ``compactness'' can be empirically obtained by measuring the gravitational redshift of pertinent spectral lines emitted from the surface of a compact star. 
Thus, from a general relativistic point of view, it is $z_{\rm b}$ and not $GM/R_{\rm b} c^2$ that is the intrinsic measure of the gravitational compactness of a star. 
The gravitational redshift  ($z_{\rm b}$) of radiation emitted from the surface of the body is given by the follwing equation~\cite{Weinberg1972,Misner1973,Thorne1977}
\begin{equation}\label{zb-eqn}
	1 + z_{\rm b}~=~\left(1 - \frac{R_{\rm S}}{R_{\rm b} }\right)^{-1/2}
\end{equation}
where $R_{\rm S} = 2 GM/c^2$ is the Schwarzschild radius of the compact object. Note that, in~the limit $GM/R_{\rm b} c^2 \ll 1$, the~foregoing equation leads to
\begin{equation}\label{New-eqn}
	z_{\rm b} \approx {GM\over R_{\rm b} c^2} = {R_{\rm S} \over 2 R_{\rm b}}
\end{equation}
and that explains why $GM/R_{\rm b} c^2$ is usually a measure of the gravitational  compactness. As~already mentioned, it is believed that an NS may have $z_{\rm b}$$\sim$$0.1$, which is almost three orders higher than that of a typical WD. However, in principle, there can be compact objects with much higher gravitational redshift ($z_{\rm b}$$\sim$$1$ or higher), and~we shall call them Ultra Compact Objects (UCOs). The central aim of this short paper is to study the effect of curved space-time around 
hot UCOs on the actual value of the luminosity of the distantly observed thermal radiation {\it vis-a-vis} ($L_{\rm \infty}^{\rm curved}$) and the luminosity 
erroneously estimated by ignoring such space-time curvature ($L_{\rm \infty}^{\rm flat}$).
\par
In general,  for~ static compact objects  that are supported by isotropic internal pressure, i.e.,~pressure is the same in both radial and transverse directions, 
$p_{\rm r} = p_{\rm \perp}$, there is a  well-known theoretical upper limit on the compactness, given by Buchdahl~\cite{Buchdahl1959} as
\begin{equation}\label{buch-eqn}
	{2G M \over R_{\rm b} c^2} = {8\over 9}
\end{equation}
and the corresponding gravitational redshift would be
\begin{equation}\label{z-Buch}
z_{\rm Buch} = 2.0
\end{equation}
It is pertinent here to note that for a static compact object of mass $M$ and radius $R_{\rm b}$, we assume that its exterior is described by the Schwarzschild 
metric possessing
\begin{equation}\label{g00-eqn}
	g_{00} = 1 - {2GM \over Rc^2}; ~~ R \ge R_{\rm b},
\end{equation}
and the gravitational redshift $z_{\rm b}$ is directly related to $g_{00}$ and attendant bending of the exterior space-time due to the effect of the 
gravity of the compact object. The~severe effect of general relativistic space-time bending around a compact object may be highlighted  by the fact 
that even light and radiation may  move in (unstable) {\em circular orbits} at $R= 1.5R_{\rm s}$ for compact stars having radii $R_{\rm b} < 1.5 R_{\rm s}$ or 
$z_{\rm b} > 0.73$. The~result follows from the studies on the propagation of photons or massless particles in  the strong gravitational field around 
sufficiently compact objects~\cite{Misner1973}. The~ closed circular orbits of photons in various planes define what is known as a \emph{photon sphere} with a 
corresponding gravitational redshift of $z_{\rm photon-sphere} = 0.73$. Note, since  $z_{\rm photon-sphere} = 0.73$  is well below the Buchdahl limit of compactness 
of $z_{\rm Buch} = 2.0$, there can be static spherical compact objects that lie within their respective photon spheres. The~most well-known compact object that  
lie within its photon sphere is the Schwarzschild black hole having $R=R_{\rm s}$ or $z_{\rm b}=\infty$. In~the following section, we shall discuss the theoretical 
possibility of having ultra-compact spherical stars that may behave like \emph{ black hole mimickers} in view of their large $z_{\rm b} \gg 1$.
%-------------------------------------------------------------------------------------------------
\subsection*{Stefan--Boltzmann Law}
From the above discussion, it is clear that for an accurate study of  emission of radiation from neutron stars and similar ultra-compact objects, the~
effect of curvature of space-time around such objects should be borne in mind. We shall highlight the fact that the faraway luminosity of such compact objects 
is higher by a factor of $(1+z_{\rm b})^2$ than what one might estimate by applying the classical Stefan--Boltzmann law of the blackbody radiation, which tells us
that a black body having a surface area $A_{\rm b}$ and temperature $T_{\rm b}$ has a thermal luminosity
\begin{equation}\label{thermal-lum}
	L_{\rm b} = \sigma ~A_{\rm b} ~T_{\rm b}^4
\end{equation}
where $\sigma \approx 5.67 \times 10^{-8}$ {\rm kg} {\rm s}$^{-3}$ {\rm K}$^{-4}$. 
\par
In 1990, Lavenda \& Dunning-Davies~\cite{Lavenda1990} argued that since black holes radiate with a thermal spectrum, they possess radiation pressure, and
Boltzmann's derivation of Stefan's Law can be applied to black holes. Since the entropy is proportional to the surface area of the black hole, %MDPI: please check to make sure intended meaning has been retained
they found that the corresponding pressure must be negative. If~the second law is not to be violated, then the temperature must also be negative. 
Since a negative temperature of the black hole seems to be self-contradictory,   these authors cast doubts on the validity of conventional black hole thermodynamics. 
Following this work linking this law with  black hole thermodynamics,  in~1995, de Lima \& Santos~\cite{Lima1995} wanted to extend this law to the 
hypothetical cosmological fluids obeying the equation of states of the form $p = (\gamma- 1) \rho/ c^2$, where $\rho$ and $p$ are the energy density and 
pressure of the fluid, respectively, and $\gamma$ is a constant. Such discussions, however, have no relevance for the present study, which concerns only 
thermal radiation $p = (1/3)  \rho/ c^2$.
\par
The Stefan--Boltzmann law concerns the thermal radiation of a black body, which is at rest with respect to the observer studying the radiation field. 
However, if the black body is moving with a velocity ${\vec v}$ with respect to the observer, one needs to extend the law by applying special relativity. 
This task was performed by Veitsman in 2013~\cite{Vietsman2013}. It was found that the observer will find most of the radiation emitted by the fast 
moving black body to be emitted parallel to the velocity vector if $ v \to c$. However, we are considering the case when there is no relative motion 
between the black body and the observer. Hence, the above study is also not applicable here. On~the other hand, even for a static observer,  in
the presence of gravity, the~space-time around the black body becomes curved, and~one must invoke general relativity to reformulate the Stefan--Boltzmann 
law and~also to interpret the result correctly. This is the aim of this~study.

%-----------------------------------Section-2:-----------------------------------------
\section{Ultra-Compact~Objects}
We have already mentioned that even a strictly static star supported by isotropic internal pressure may possess $z_{\rm b} \approx 2.0$ and, accordingly, 
lie well within its photon sphere ($z_{\rm photon-sphere} = 0.73)$.  Now, we recall that if the compact star is supported by anisotropic internal 
pressure ($p_{\rm r} \neq p_\perp$), its compactness may exceed the Buchdahl upper limit. Following the landmark paper by Buchdahl~\cite{Buchdahl1959}, 
the topic of study of self-gravitating static anisotropic spheres has  been carried out by many authors by considering various restrictions on the 
degree of anisotropy, different  equation of states and several latent physical conditions. One important point is that incorporation of pressure 
anisotropy not only raises the upper limit on $z_{\rm b}$, but also the upper limit on the maximum mass of the compact stars. Nonetheless, here, we are primarily 
interested in the former aspect alone. 
\par
Two of the significant initial studies in this direction were by Bondi (1964) \cite{Bondi1964} and Bowers \& Liang (1974) \cite{Bowers1974}. Additionally, two recent 
general studies on this topic are due to Herrera, Ospino \& Prisco (2008) \cite{Herrera2008} and Herrera \&  Barreto (2013) \cite{Herrera2013}.
However, since we are interested here in the numerical value of $z_{\rm b}$, we may note an earlier detailed study by Ivanov (2002) \cite{Ivanov2002}, which 
shows that the upper limit on the gravitational redshift for an anisotropic static spherical star could be bounded as $z_{\rm b} < 5.211$ . Later, as B\"ohmer (2006)~\cite{Bohmer2006} studied the problem independently, he too found an upper limit of $z_{\rm b} < 5.0$, which is more or less the same as what Ivanov (2002) had 
found earlier. 
\par
The above mentioned upper limits on gravitational redshift for anisotropic compact stars may not be the final word on this issue. A~more recent study, on~the 
possibilty that some of the black hole candidates might be ultra-compact objects, claims that for some choices of model parameters, an~anisotropic star could 
not only be arbitrarily compact ($z_{\rm b} \gg 1$), but~could be as massive as many black holes too~\cite{Guilhermeet2019}.
%--------------------------------------------------------------------------------------------
\subsection*{Quasi-Static Ultra-Compact Objects}
In general relativity, as~already mentioned, the~exterior space-time of a {\em strictly  static} spherically symmetric self-gravitating object is given by 
the {\em vacuum} Schwarzschild exterior metric. Thus, in~a strict sense, the~exterior space-time of a radiating star is non-static, even though the degree 
of non-staticity may be  negligible. On~the other hand, the~Buchdahl upper limit ($z_{\rm b} < 2.0$) is  obtained by considering a strictly static 
non-radiating exterior space-time. Therefore, in~principle, the~surface gravitational redshift of a radiating star may exceed the Buchdahl limit, even if 
one would ignore any anisotropy due to inhomogeneity or magnetic field. It was found that, accordingly, there could be radiating quasi-static ultra-compact objects having $z_{\rm b} \gg 1$, which are supported against their intense self-gravity by the outward pressure of the radiation trapped by the 
same intense self-gravity~\cite{Mitra2006a,Mitra2006b,Mitra2006c,Mitra2010}.

%------------------------------------Section-3:-------------------------------------------------------
\section{Effects of Gravitational~Redshift}
The physical interpretation of gravitational redshift is that an amount of energy $dE_{\rm b}$ on the surface of a black body will be measured as~\cite{Weinberg1972,Misner1973,Thorne1977}
\begin{equation}
	dE_{\rm \infty} = \frac{dE_{\rm b}}{1+z_{\rm b}}
\end{equation}
by an infinitely faraway observer. Additionally, if an interval of time measured on the surface of the black body is $dt$, the~same will become dilated to~\cite{Weinberg1972,Misner1973}
\begin{equation}
	dt_{\rm \infty} = (1+z_{\rm b})~dt
\end{equation}
to the same faraway observer.
\par
Accordingly, the~luminosity measured by the faraway observer is
\begin{equation}
	\frac{dE_{\rm \infty}}{dt_{\rm \infty}} = (1+z_{\rm b})^{-2} ~\frac{dE_{\rm b}}{dt}.
\end{equation}
Therefore, the~ faraway luminosity will be smaller than the locally measured luminosity by a factor of $(1+z_{\rm b})^{-2}$, irrespective of the physical 
mechanism of the radiation, i.e, thermal or non-thermal~\cite{Thorne1977}:
\begin{equation}\label{Linf}
	L_{\rm \infty} = (1+z_{\rm b})^{-2} ~L_{\rm b}
\end{equation}%MDPI: Please check if the dot should be deleted.

%----------------------------------------------------------------------------------------------------------------
\subsection*{UCOs Lie within Their Photon Spheres}
The foregoing equation implicitly assumes that all the radiation quanta leaving the compact object reach up to the faraway observer and~do not return towards the 
compact object.  Suppose the~emission of the radiation is taking place from a given patch or few patches on the surface of the compact object and not from the entire 
surface; and in such a case, {\em the assumption of spherical symmetry will become violated}. Even for such anisotropic cases, as~long as the compact object is larger than 
its photon sphere, $R_{\rm b} \ge R_{\rm photon-sphere} = 1.5 R_{\rm S}$, no radiation quanta bends backwards during their outward journey irrespective of their direction 
of emission, i.e.,~ even if they would be emitted in a non-radial direction. 
\par
Does Equation~(\ref{Linf}) change for compact objects lying within their respective photon spheres$?$ In~order to probe, this we recall that all realistic gravitational 
collapse processes involve emission of radiation~\cite{Mitra2006a}, and UCOs are formed by the emission of radiation. When the compact object lies within its photon sphere, 
$R_{\rm b} < 1.5 R_{\rm S}$, the~quanta emitted in the non-radial directions may bend backwards by the effect of strong gravity, and~only the radiation emitted within a 
cone defined by a semi-angle $\theta_{\rm c}$ \cite{Misner1973}
\begin{equation}
	\sin \theta_{\rm c} = {\sqrt{27}\over 2} (1- R_{\rm S}/R_{\rm b})^{1/2} (R_{\rm S}/R_{\rm b})
\end{equation}
can move away to infinity.
\par
If $R_{\rm b} \approx R_{\rm S}$, the~foregoing equation reduces to
\begin{equation}
	\sin \theta_{\rm c} \to \theta_{\rm c} \approx  {\sqrt{27}\over 2} (1+z_{\rm b})^{-1}.
\end{equation}
In~such a case, the~solid angle of the radiation that can travel to infinity becomes
\begin{equation} 
	\Omega_{\rm c} \approx \pi \theta_{\rm c}^2 = {27 \pi  \over 4} (1+z_{\rm b})^{-2}
\end{equation}
\hl{Accordingly}, if~the {\em radiation from the given patch} is emitted in a solid angle of  $2 \pi$ and not exclusively in a radially outward direction, the~chance of escape 
of radiation from the given patch would decrease by a factor of $\Omega_{\rm c} /2 \pi \approx (27/8) (1+z_{\rm b})^{-2}$. In~fact, the radiation emitted within the body 
of the compact object always interacts with trillions of atoms and electrons and~moves only diffusively by performing random walk. As~a result, once a compact object 
dips within its photon sphere, the~radiation in the interior of the body may become gravitationally trapped~\cite{Mitra2006b,Mitra2006c,Mitra2006d,Mitra2010}. However,  the  
radiation quanta emitted from the surface of the compact object into the exterior {\em vacuum}  moves unhindred without any random walk. In~particular, {\em if a  quanta 
is ejected radially outward along the local normal direction} $\theta_{\rm c} =0$, then {\em it can continue to move radially outward howsoever large $z_{\rm b}$ might be}. 
Thus, if~the compact object is emitting radiation truly {\em isotropically in a radially outward direction at each point of emission}, all the  outwardly moving quanta 
reach infinity even if $z_{\rm b} \gg 1$. In~such a case, Equation~(\ref{Linf}) connecting local luminosity and distantly measured luminosity  will remain valid 
irrespective of the value of $z_{\rm b}$. This important fact can be confirmed by noting the pertinent equation for the continued spherical gravitational contraction 
of stars~\cite{Misner1965,Misner1973,Hernandez1966,Herrera2004,Herrera2006,Tewari2013}%MDPI: Please check if an indent should be added. Please check throughout the main text and add the indent if necessay.
\begin{equation}\label{L-collapse}
	L_{\rm \infty} = (U_{\rm b} + \Gamma_{\rm b})^2 L_{\rm b}
\end{equation}
where the parameter
\begin{equation}
	U_{\rm b} =  {\partial R_{\rm b} \over \partial \tau}
\end{equation} 
is the rate of contraction of the outer radius of the star, with respect to local comoving proper time $\tau$.
On the other hand, the~other parameter is
\begin{equation}
	\Gamma_{\rm b} = \left(1 +U_{\rm b}^2 - \frac{2GM}{R_{\rm b} c^2}\right)^{1/2}
\end{equation}
(see, e.g.,~Equation~(5.10) in~\cite{Misner1965}, Equation~(37) in~\cite{Hernandez1966}, Equation~(18) in~\cite{Herrera2004}, \linebreak  \mbox{Equation~(20)} in~\cite{Herrera2006} or 
\mbox{Equation~(17)} in~\cite{Tewari2013}). Now suppose that the continued gravitational collapse has resulted in the formation of a static UCO ($U_{\rm b}=0$) or quasi-static 
UCO ($U_{\rm b} \to 0$). In~such a case, one will have
\begin{equation}
	\Gamma_{\rm b} = \left(1 - \frac{2GM}{R_{\rm b} c^2}\right)^{1/2} = (1+z_{\rm b})^{-1},
\end{equation}
and Equation~(\ref{L-collapse}), connecting local and distant luminosity, will reduce to
\begin{equation}\label{L-obs}
	L_{\rm \infty} = (1+z_{\rm b})^{-2}~ L_{\rm b}.
\end{equation} 
Since Equation~(\ref{L-collapse}) is valid all the way up to $R_{\rm b} \ge R_{\rm S} = 2GM/c^2$ and not merely up to $R_{\rm b} \ge 1.5 R_{\rm S} = 3 GM/c^2$, for
a spherically symmetrical compact object radiating isotropically along the local outward normal direction, Equation~(\ref{L-obs}) is valid all the way up to 
$z_{\rm b} \to \infty$. Therefore,  as~before, the faraway luminosity of a compact object lying well within its photon sphere  will be smaller than the locally 
measured luminosity by a factor of $(1+z_{\rm b})^{-2}$, irrespective of the physical mechanism behind the radiation. The~fact that  a black hole is not visible to any 
faraway observer  may also be physically understood by noting that $(1+z_{\rm b})^{-2}=0$ for $z_{\rm b} = \infty$.
%----------------------------------------------------------------------------------------
\section{Distant Luminosity of Thermal~Radiation}
It is now clear that if  the surface temperature of a spherical black body is $T_{\rm b}$, the~faraway thermal luminosity will accordingly be
\begin{equation}
	L_{\rm \infty} = (1+z_{\rm b})^{-2}~ L_{\rm b} = (1+z_{\rm b})^{-2}~ \sigma A_{\rm b} T_{\rm b}^4
\end{equation}
irrespective of the value of $z_{\rm b}$. From~the foregoing equation, one may be tempted to conclude that the temperature of the black body as perceived 
by a faraway observer would~be
\begin{equation}\label{Tinfinite-eqn}
	T_{\rm \infty} = \left(\frac{L_{\rm \infty}}{\sigma A_{\rm b}}\right)^{1/4} =  \frac{T_{\rm b}}{\sqrt{1+z_{\rm b}}} ~ \rm (incorrect).
\end{equation}
 Now we come to the crucial aspect of this paper and expain below that the foregoing  conclusion would be incorrect.
\par
In the curved space-time, {\em the temperature of the blackbody radiation varies spatially} in a manner that was discovered by 
Tolman \& Ehrenfest in 1930~\cite{Tolman1930a,Tolman1930b,Tolman1934}, 14 years after the formulation of general relativity and 50 years after the formulation 
of the Stefan--Boltzmann law. This important problem was extended for any stationary space-time by Buchdahl in 1949~\cite{Buchdahl1949}. It follows that 
in the curved space-time around the compact object, the~temperature varies following the rule
\begin{equation}\label{Tg00-eqn}
T \sqrt{|g_{\rm 00}|} = \rm constant.
\end{equation}
From Equations~(\ref{zb-eqn}) \& (\ref{g00-eqn}), since $g_{\rm 00} = 1$ at $R=\infty$, one finds that  the temperature of the black body radiation at  
$R=\infty$ is lower by a factor of $(1+z_{\rm b})$:
\begin{equation}\label{Tinf}
	T_{\rm \infty} =  \frac{T_{\rm b}}{1+z_{\rm b}}.
\end{equation}
By combining Equations~(\ref{Tinfinite-eqn}) and (\ref{Tinf}), it is found that in terms of the faraway temperature $T_{\rm \infty}$, the~faraway luminosity 
is
\begin{equation}\label{Linf-new}
	L_{\rm \infty} = (1+z_{\rm b})^{2} ~ \sigma A_{\rm b} T_{\rm \infty}^4.
\end{equation}
However, if~the effect of gravitation around the compact object would be ignored, the~exterior space-time would be flat with no gravitational redshift at 
all ($z_{\rm b}=0$), and~one would~have
\begin{equation}
	T_{\rm \infty}^{\rm flat} = T_{\rm b}
\end{equation}
and accordingly
\begin{equation}
	L_{\rm \infty}^{\rm flat} = L_{\rm b} = \sigma A_{\rm b} T_{\rm b}^4.
\end{equation}
These two equations lead us to our conclusion that for a spherically symmetric black body emitting radiation isotropically along local normal directions, 
one has
\begin{equation}
	L_{\rm \infty}^{\rm curved} = (1+z_{\rm b})^2 ~ L_{\rm \infty}^{\rm flat}.
\end{equation} 
%---------------------------------------------------------------------------------------------------
\subsection*{Does the Radius of the Blackbody change by Gravity$?$}
Suppose the faraway observer  is unmindful of the general relativistic effect described here and chooses to define the thermal luminosity recorded 
by him/her as
\begin{equation}
	L_{\rm \infty} = 4 \pi \sigma R_{\rm bb}^2 T_{\rm \infty}^4.
\end{equation}
Since $A_{\rm b} = 4 \pi R_{\rm b}^2$, then by comparing the above equation with Equation~(\ref{Linf-new}), he/she might conclude that the~effect of 
gravity has increased the black body radius as
\begin{equation}
	R_{\rm bb} = (1+z_{\rm b}) R_{\rm b}.
\end{equation}
However,~let us explain why the above interpretation about the enhancement of radius is incorrect. Note that in a spherically symmetric space-time, surface area 
of a sphere $A= 4 \pi R^2$ is defined covariantly~\cite{Einstein2016}, and~thus, the enhancement of the luminosity over the corresponding flat space-time 
must not be interpreted in terms of an enhanced effective radius
\begin{equation}
	R_{\rm \infty} = R_{\rm bb} = (1+z_{\rm b}) R_{\rm b} ~~ \rm (incorrect).
\end{equation}
It will be important here to recall that for a positively curved homogeneous space-time described by the Friedmann metric, while the proper volume of a 
sphere $V_{\rm GR} = 2 \pi^2 R^3$ differs significantly from the Euclidean formula $V_{Euc} = (4\pi/3) R^3$, the~surface area of a sphere of radius $R$ continues 
to be unchanged $4\pi R^2$ \cite{Hypertext22,Misner1973}. One may also appreciate the fact that the surface area of a sphere $A= 4 \pi R^2$ is an invariant by noting 
that for all spherically symmetric space-time metrics, be it Euclidean or non-Euclidean, the~angular part of the metric always contain the term 
$R^2 d\Omega^2$, where $d\Omega^2$ is the metric of a unit sphere.
\par
Thus, the radius of a spherical blackbody does not differ for two observers because {\em area of a sphere is an invariant even for a curved space-time}.
In fact, it is because this invariance of the area of a spherical surface that the~radial coordinate $R$ appears as the ``luminosity distance'' even for a (static) 
curved~space-time. %MDPI: please check to make sure intended meaning has been retained.

%--------------------------------------Section 5----------------------------------------------------
\section{Discussions and~Conclusions}
For the curved space-time around a hot compact object having a surface gravitational redshift of $z_{\rm b}$, the~actual luminosity infinitely faraway from 
the source is higher by a factor of $(1+z_{\rm b})^2$ than what one may naively estimate by a tacit flat space-time computation by  using the  far away local 
temperature of the radiation field (Equations (23) and (26)). The~effect described here is absent or negligble or not significant in most of the standard 
astrophysical discussions probably~because:
\begin{itemize}
\item Most of the radiation high energy that astrophysicists consider are {\em non-thermal} while the present discussion is pertinent only for {\em thermal radiation 
	emitted spherically from the surface} of a compact~object.

\item For WDs, compactness or $z_{\rm b} \sim 10^{-4}$ is extremely low, and~therefore, this effect can be safely~ignored.

\item For hot NSs, this effect is indeed relevant for  thermal emission  due to a hot surface or Type I X-ray bursts, with thermal flashes resulting 
	from  runaway thermonuclear burning of accreted matter~\cite{ParLewin1988}. However, since that is usually believed for NSs,  
	$z_{\rm b} \sim 0.1$, and~this effect is very significant: $L_{\rm \infty}^{\rm curved} = 1.21 ~ L_{\rm \infty}^{\rm flat}$.
	
\item For X-ray binaries containing BHs, the~thermal X-ray emission mostly originates from extended hot accretion disks and not from the surface of the BHs. 
Thus, this effect is not pertinent for thermal X-ray emission from BH X-ray binaries.
\end{itemize}
\par
A closer inspection, however, shows that for a canonical NS having $R= 10$ km and $M = 1.4$ solar-mass, actually, $z_{\rm b} \approx 0.15$, and~this effect may be important 
because the distantly observed thermal luminosity will be larger than the flat space-time luminosity by a factor of $1.32$ or $32 \%$ higher.
\par
As already stressed, the~enhancement factor of the distant luminosity remains unaltered even when the compact object lies within its photon sphere $z_{\rm b} > 0.73$ because 
for truly isotropic emission, at~each point, radiation quanta is emitted along the local normal or radial direction: $\sin \theta_{\rm c}=0$.
\par
Accordingly, this effect could be highly significant for UCOs and Black Hole Mimickers, which are theoretically admissible compact objects 
with $z_{\rm b} \sim 1$ or even $z_{\rm b} \gg 1$ \cite{Guilhermeet2019,Mitra2006a,Mitra2006b,Mitra2010}.
\par
If we consider the theoretical upper limit of $z_{\rm b} <5.211$ \cite{Ivanov2002,Bohmer2006} of static UCOs, the~actual faraway thermal luminosity could be nearly 36 times 
higher than the flat space-time case. Even if we consider the Buchdahl upper limit of $z_{\rm b} <2.0$, the~actual thermal luminosity could still be nine times higher than the 
corresponding flat space-time case.

%%%%%%%%%%%%%%%%%%%%%%%%%%%%%%%%%%%%%%%%%%
\section*{Author Contributions}
Conceptualization and  writing---original draft preparation, A.M.; validation, writing---review and editing, K.K.S. All authors have read and agreed to the published 
version of the manuscript.

%%%%%%%%%%%%%%%%%%%%%%%%%%%%%%%%%%%%%%%%%%
\section*{Funding}
This research received no external~funding.

%%%%%%%%%%%%%%%%%%%%%%%%%%%%%%%%%%%%%%%%%%
\section*{Acknowledgments}
Authors thank all the four anonymous referees for their important comments and suggestions which led to the improvement of the contents of the~manuscript.

%%%%%%%%%%%%%%%%%%%%%%%%%%%%%%%%%%%%%%%%%%
\section*{Conflicts of Interest}
The authors declare no conflict of~interest. 

%=====================================
% References, variant A: internal bibliography
%=====================================

\end{document}